\documentclass[prl,superscriptaddress,showpacs,keywords,reprint]{revtex4-1}
\usepackage{graphicx,color}
\usepackage{amsfonts,amsmath, amsthm, amssymb,graphicx,hyperref}

\newcommand{\be}{\begin{equation}}

\newcommand{\ee}{\end{equation}}
\newcommand{\ben}{\begin{eqnarray}}
\newcommand{\een}{\end{eqnarray}}
\newcommand{\bes}{\begin{subequations}}
\newcommand{\ees}{\end{subequations}}
\newcommand{\bF}{\begin{figure}}
\newcommand{\eF}{\end{figure}}

\def\tr{ {\rm{Tr }}\,}

\newcommand{\kt}{\rangle}
\newcommand{\br}{\langle}

\usepackage{lineno,hyperref}
\modulolinenumbers[5]

\begin{document}
 \title{Quantum chaos, thermalization and tunneling in an exactly solvable few body system}

\author{Shruti Dogra}
\affiliation{Department of Physics, Indian Institute of Technology Madras, Chennai, India 600036}

\author{Vaibhav Madhok}
\affiliation{Department of Physics, Indian Institute of Technology Madras, Chennai, India 600036}

\author{Arul Lakshminarayan}
\affiliation{Department of Physics, Indian Institute of Technology Madras, Chennai, India 600036}

\begin{abstract}
Exactly solvable models that exhibit quantum signatures of classical chaos are both rare as well as important - more so in view of the fact that the mechanisms for ergodic behavior and thermalization in isolated quantum systems and its connections to non-integrability are under active investigation. In this work, we study quantum systems of few qubits collectively modeled as a kicked top, a textbook example of quantum chaos. In particular,
 we show that the 3 and 4 qubit cases are exactly solvable and yet, interestingly, can display signatures of ergodicity and thermalization.
Deriving analytical expressions for entanglement entropy and concurrence, we see agreement in certain parameter regimes between long-time average values and ensemble averages of random states with permutation symmetry. Comparing with results using the data of a recent transmons based experiment realizing the 3-qubit case, we find agreement for short times, including a peculiar step-like behaviour in correlations of some states. In the case of 4-qubits we point to a precursor of dynamical tunneling between what
in the classical limit would be two stable islands. Numerical results for larger number of qubits show the emergence of the classical limit including signatures of a bifurcation.
 
\end{abstract}

\maketitle


In a modest pursuit of the esthetic attributed to Feller that ``the best consists of the general embodied in the concrete" \cite{Billingsley}, we consider extreme
quantum cases of the kicked top, a widely studied, text-book model of quantum chaos \cite{KusScharfHaake1987,Haake,Peres02}. The general issues at hand are the emergence
of the classical world from a quantum substratum and the role of quantum chaos in the thermodynamics of closed quantum systems \cite{CassidyEtal2009,SantosRigol2010,Rigol16}. Vigorous progress is being made in studying thermalization of isolated quantum systems that could be either time-independent or periodically forced \cite{JensenShankar1985,Deutsch91,Srednicki94,Rigol2009,CassidyEtal2009,SantosRigol2010,CiracHastings2011,
DeutchLiSharma2013,LangenEtal2013,Rigol16,LucaRigol2014,
LazaDasMoess2014,LazDasMoess2014pre,Haldar2018,Neill16,Kaufman2016,ClosEtal2016,HazzardEtal2014}. 
  Entanglement within many-body quantum states in such systems 
drives subsystems to thermalization although the full
state remain pure and of zero entropy, see \cite{Kaufman2016} for a demonstration with cold atoms.

Quantum chaos \cite{Gutzwiller1990,Haake} and, consequently, eigenstate thermalization hypothesis \cite{Srednicki94,Rigol16} enables one to use individual states for ensemble averages. For periodically driven systems that do not even conserve energy, a structureless ``infinite-temperature" ensemble emerges in strongly non-integrable regimes \cite{LucaRigol2014,LazDasMoess2014pre}. 
A recent 3-qubit experiment, using superconducting Josephson junctions, that simulated the kicked top \cite{Neill16} (see also \cite{Madhok2018_corr}) purported to remarkably demonstrate such a thermalization. Although such behavior has been attributed to non-integrability \cite{Neill16,Rigol16}, we exactly solve this 3-qubit kicked top, pointing out that it can be interpreted as a special case of an {\it integrable} model. Additionally we  solve the 4-qubit case exactly, although there is no connection to an integrable model.  
The kicked top, in the limit of an infinite number of qubits displays a standard transition to Hamiltonian chaos and it is remarkable that many of the features are already reflected in the solvable few qubit cases. 

For example, explicit formulas are obtained for entanglements generated and are compared, for the 3-qubit case, with data from the experiment in \cite{Neill16}. The infinite time average of single qubit entanglement is found analytically for some initial states and at a special and 
large value of the forcing, for all initially unentangled coherent states. These are shown to tend to that obtained from relevant (random matrix) 
ensembles, in some cases even exactly coinciding with them and  thus displaying thermalization. These demonstrate that even in the deep quantum regime, the transition to what in the classical limit becomes chaos is reflected in the time-averaged entanglement. 

Naturally there are interesting quantum effects in these few-body systems. One, is the extremely slow convergence of subsystem entropies in the near-integrable regime that happens for some states of the 4-qubit case. Its origin is the presence of dynamical tunneling 
\cite{DavisHeller1981,LinBallentine1990,Peres1991,Tomsovic98b,SrihariBook} 
between what appears in the classical limit as symmetric regular regions. 
In the near-integrable regime the exactly calculable tunneling splitting is shown to result in this long-time dynamics. This may open windows for 
experimental tests of the interplay of chaos, resonances and tunneling in systems with small number of qubits.

 
 
The quantum kicked top is a combination of rotations and torsions, the Hamiltonian \cite{KusScharfHaake1987,Haake,Peres02} is given by
$H=(\kappa_0/2j){J_z}^2 \sum_{n = -\infty}^{ \infty} \delta(t-n\tau)+(p/\tau) \, {J_y}.$
Here $J_{x,y,z}$ are components of the angular momentum operator $\mathbf{J}$. 
The time between periodic kicks is $\tau$. The Floquet map is the unitary operator
$\mathcal{U} = \exp\left [-i (\kappa_0/2j \hbar) J_z^2 \right]\exp\left[-i (p/\hbar) J_y\right]$,
which evolves states just after a kick to just after the next. The parameter $p$ measures rotation about the $y$ axis, and in the following we set $\hbar=1$ and $p=\pi/2$.
$\kappa_0$ is the magnitude of a twist applied between kicks and controls the degree of chaos in 
the classical system. As the magnitude of the total angular momentum is conserved, the quantum number $j$, with eigenvalues of $\mathbf{J}^2$ being $j(j+1)\hbar^2$,  is a good one. The classical limit, when $j \rightarrow \infty$ is a map of the unit sphere phase space $X^2+Y^2+Z^2=1$ onto itself with the variables being $X,Y,Z=J_{x,y,z}/j$ and is given by $(X'=Z\cos(\kappa_0 X)+Y\sin (\kappa_0 X),\,Y'=-Z\sin(\kappa_0 X)+Y\cos (\kappa_0 X),\, Z'=-X)$. 

For $\kappa_0=0$ the classical map is evidently integrable, being just a rotation, but for $\kappa_0>0$ chaotic orbits appear in the phase space and when  $\kappa_0>6$ it is essentially fully chaotic. Connection to a many-body model can be made by considering the large $\mathbf{J}$ spin as the total spin 
of spin=1/2 qubits, replacing $J_{x,y,z}$ with $\sum_{l=1}^{2j} \sigma^{x,y,z}_l/2$ \cite{Milburn99,Wang2004}. The Floquet operator is then that of $2j$ qubits, an Ising model with all-to-all homogeneous coupling and a transverse magnetic field:
\begin{equation}
\label{uni}
{\mathcal U}=\exp\left(-i \frac{\kappa_0}{4j}  \sum_{ l< l'=1}^{2j} \sigma^z_{l} \sigma^z_{l'}\right)
\exp\left( -i \frac{\pi}{4} \sum_{l=1}^{2j}\sigma^y_l \right).
\end{equation}
Here $\sigma^{x,y,z}_l$ are the standard Pauli matrices, and an overall phase is neglected. The case of 2-qubits, $j=1$, has been analyzed in \cite{RuebeckArjendu2017} wherein interesting arguments have been proposed for the observation of structures not linked to the classical limit.
For $j=3/2$, the three qubit case is a nearest neighbor kicked transverse Ising model, known to be integrable \cite{Prosen2000,ArulSub2005}. For higher values of the spin, the model maybe considered few-body realizations of non-integrable systems. In general only the $2j+1$ dimensional permutation symmetric subspace of the full $2^{2j}$ dimensional space is relevant to the kicked top.

\begin{figure}
 \centering
 \includegraphics[scale=1,keepaspectratio=true]{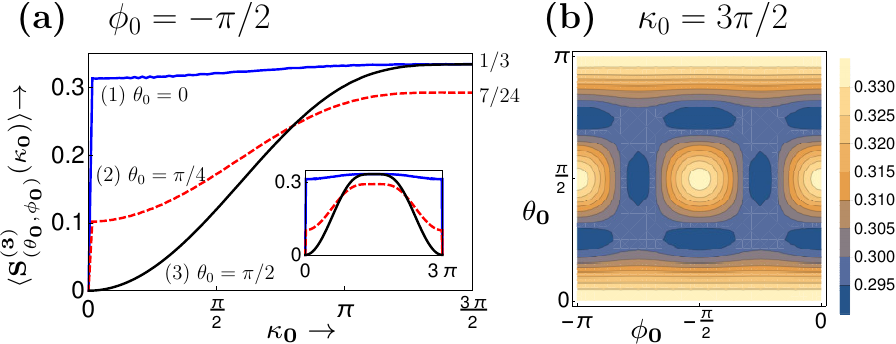}
 \caption{(a) Time averaged linear entropy, obtained over $n=1000$ periods,
 of a single qubit {\it vs} the parameter $\kappa_0$, for three initial coherent states $|\theta_0, \phi_0\kt$. The Eqs.~(\ref{eq:avgpiby2}) apply to the curves labeled (1) and (3), as for $\theta_0=0$ the value of $\phi_0$ is immaterial on the sphere.  Inset shows the entanglement periodicity in the parameter at $\kappa_0=3\pi$. Part (b) displays the time averaged linear entropy across all initial coherent states for the value $\kappa_0=3\pi/2$ and is described by Eq.~(\ref{eq:entthetaphi}).}
\label{fig:entropyplot}
\end{figure}

The initial states used are coherent states located at $(X=\sin\theta_0 \cos\phi_0, \,Y= \sin \theta_0 \sin\phi_0, \, Z=\cos \theta_0)$ on the phase space sphere and given by $|\theta_0,\phi_0\kt = \otimes^{2j} (\cos(\theta_0/2) |0\kt + e^{-i \phi_0} \sin(\theta_0/2) |1\kt)$ \cite{Glauber,Puri}. 
Note that for $\kappa_0$ that are multiples of $2 \pi j$, $\mathcal{U}$ is a local operator and does not create entanglement, we therefore restrict attention to the interval $\kappa_0 \in [0, \pi j]$. After time $n$ the evolved state $|\psi_n\kt =\mathcal{U}^n|\theta_0,\phi_0\kt$ is used
to find the reduced density matrix $\rho_1(n)=\tr_{\neq 1}(|\psi_n \kt \br \psi_n |)$, obtained after tracing out all other spins except the first.
As this is at most rank-2, the various entropies that depend on the eigenvalues alone are monotonic to each other and we use the simplest, the linear entropy $S^{(2j)}_{(\theta_0,\phi_0)}(n,\kappa_0)= 1-\tr \rho_1^2(n)$ as a measure of entanglement.
Figure~(\ref{fig:entropyplot}) shows the long-time average  $\br S^{(3)}_{(\theta_0,\phi_0)}(\kappa_0)\kt$ as a function of $\kappa_0$, for the case of 3-qubits, and three representative initial states. Each is seen to increase with the torsion $\kappa_0$ either to $1/3$ or a value close to it. 

The average value of the linear entropy in the $N$-qubit permutation symmetric subspace is $S_{RMT}(N)=(N-1)/(2N)$  \cite{2018arXiv180600113S}, and for $N=3$ this also gives $1/3$. For at least three particular initial states, with important classical phase space correspondences, $|0,0\kt\equiv |000\kt$ and $|\pi/2, \pm \pi/2\kt \equiv |\pm\pm\pm\kt_y$ this value is, remarkably, exactly attained for $\kappa_0=3\pi/2$. For these states 
\begin{equation}
\label{eq:avgpiby2}
\br S^{(3)}_{(0,0)}(\kappa_0)\kt=\frac{5-2 s_0}{\left(4-s_0\right)^2},\,\br S^{(3)}_{(\frac{\pi}{2},\pm \frac{\pi}{2})}(\kappa_0)\kt=s_0\frac{ 8-5 s_0}{\left(4-s_0 \right)^2},
\end{equation}
with $\, s_0=\sin^2\left(\kappa_0/3\right)$ and $\kappa_0>0$ (see below and \cite{SM}).
While $j=3/2$ is too small to see effects such as the fixed points' loss of stability, the overall region surrounding the classical fixed points $(\theta_0,\phi_0)=(\pi/2, \pm \pi/2)$ being stable for small $\kappa_0$ and gradually losing stability as the parameter is increased is reflected in the gradual increase of average entropy corresponding to the
initial states $|\pi/2,\pm \pi/2\kt$ starting from $0$ when $\kappa_0=0$. Notice that from a purely quantum mechanical view, $\otimes^{2j} |\pm \kt_y$ are eigenstates of $\mathcal{U}$ at $\kappa_0=0$.  In contrast, the initial state  $|0,0 \kt$ corresponds to a classical period-4 orbit and assumes entanglement entropy as large as $5/16$ for arbitrarily small $\kappa_0$. 
  
For the 3-qubit case, when $\kappa_0=3\pi/2$, the eigenvalues of $\mathcal{U}$
are $\exp(\pm 2 \pi i/3)$ and $\pm \exp(\pm \pi i/6)$, implying that $\mathcal{U}^{12}=I$. Thus infinite 
time averages are finite ones over a period, in fact entanglement has a period of $6$ in this case and 
for arbitrary initial coherent states, the time-averaged entanglement entropy is
\begin{equation}
\begin{split}
\br S^{(3)}_{(\theta_0,\phi_0)}(3\pi/2)\kt = &\frac{1}{48}[ 15+\cos(4 \theta_0) + \\ &(1+3 \cos(2 \theta_0)) \sin^4 \theta_0 \sin^2(2 \phi_0)].
\end{split}
\label{eq:entthetaphi}
\end{equation}
This takes values in the narrow interval $[7/24,1/3]$, and is shown in Fig.~(\ref{fig:entropyplot}). The minimum corresponds to several initial states including $|\pi/4,\pm \pi/2\kt$ and the maximum includes the $|0,0\kt$ and $|\pi/2, \pm \pi/2\kt$ states as already noted above. The structures seen are not directly linked to classical phase space orbits, except through shared symmetries \cite{RuebeckArjendu2017}, and cannot be expected to do so as the classical limit is for fixed $\kappa_0$ and $j \rightarrow \infty$. 
Nevertheless these results lend quantitative credence to thermalization in the sense that the time averaged entropy of subsystems of most states are close to the ensemble average for suitable large $\kappa_0$, even for the 3-qubit case \cite{Neill16,Rigol16}.

The solution to the 3-qubit case proceeds from the general observation that $[\mathcal{U},\otimes_{l=1}^{2j} \sigma^y_l]=0$, {\it i.e.,} there is
an ``up-down" or parity symmetry.
The standard 4-dimensional spin quartet permutation symmetric space with $j=3/2$, $\{|000\kt, |W\kt=(|001\kt+|010\kt+|100\kt)/\sqrt{3},
|\overline{W}\kt =(|110\kt+|101\kt+|011\kt)/\sqrt{3},|111\kt\}$ is parity symmetry adapted to form the basis
\begin{equation}
\left \{|\phi^{\pm}_1\kt=\frac{1}{\sqrt{2}}(|000\kt \mp i | 111 \kt), \, |\phi_2^{\pm}\kt=\frac{1}{\sqrt{2}} (|W\rangle \pm i |\overline{W}\kt) \right \}.
\end{equation}
In this basis $\mathcal{U}$ block diagonalizes into two $2\times 2$ blocks,
\begin{equation}
  \label{eq:Uplusm}
  \mathcal{U}_{\pm} = \pm  e^{\mp \frac{i \pi}{4}} e^{-i \kappa} \begin{pmatrix}
   \frac{i}{2}e^{-2i \kappa} & \mp \frac{\sqrt{3} }{2} e^{-2i \kappa} \\
   \pm \frac{\sqrt{3}}{2} e^{2i \kappa} &  -\frac{i}{2}e^{2i \kappa}
   \end{pmatrix},
 \end{equation}
  corresponding to parity eigenvalue $\pm1$, spanned by 
$\{|\phi^+_1 \kt, |\phi^+_2\kt \}$, and  $\{|\phi^-_1 \kt, |\phi^-_2\kt \}$. For simplicity the parameter $\kappa=\kappa_0/6$ is used in these expressions. To evolve initial states we need $\mathcal{U}^n$ and therefore
$\mathcal{U}_{\pm}^n$. Expressing Eq.~(\ref{eq:Uplusm}) as a rotation and a phase, enables the explicit formula \cite{SM}
\begin{equation}
\label{eq:Upluspowern}
\mathcal{U}_{\pm}^n = (\pm 1)^n e^{-i n (\pm  \frac{\pi}{4}+\kappa)}
\begin{pmatrix}
 \alpha_n &
   \mp \beta_n^* \\
   \pm \beta_n &  
   \alpha_n^*
 \end{pmatrix}, 
\end{equation}   
   where $\alpha_n = T_n(\chi)+\frac{i}{2}\, U_{n-1}(\chi) \cos 2\kappa $ and 
 $ \beta_{n} = (\sqrt{3}/2)\, U_{n-1}(\chi) \,e^{2i \kappa}$. The Chebyshev polynomials $T_n(\chi)$ and $U_{n-1}(\chi)$ are defined as $T_n(\chi)=\cos(n \theta)$ and $U_{n-1}(\chi)=\sin(n \theta)/\sin \theta$ with 
   $\chi=\cos{\theta}=\sin(2\kappa)/2$.

\begin{figure}
 \centering
 \includegraphics[scale=1]{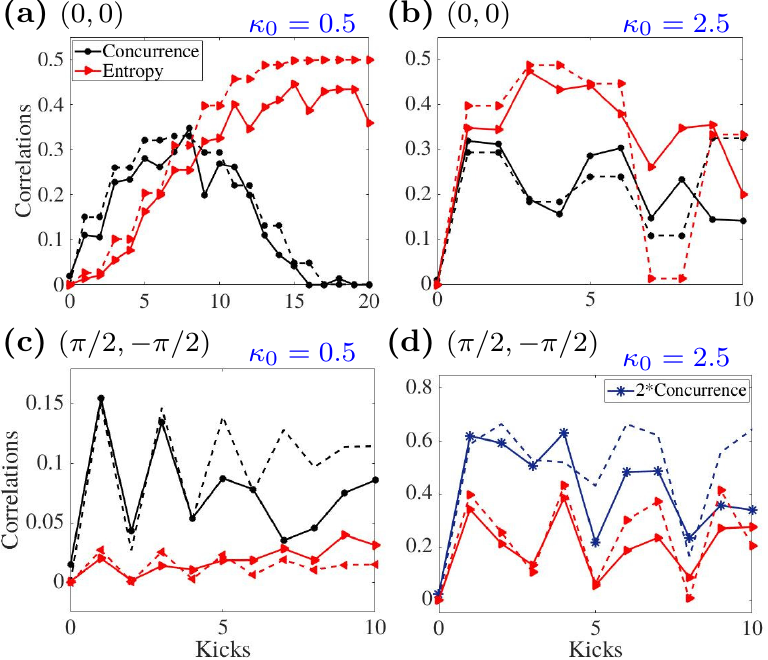}
 \caption{Plots showing analytical (dashed curves with markers), experimental 
 (solid curves with markers) and numerical (dashed) curves of linear entropy 
 and concurrence as a function of the number of kicks, as the initial state $|\psi_0\rangle$ is evolved  
  under repeated applications of operator $\mathcal{U}$.
 Parameters of the initial state, $(\theta_0, \phi_0)$, and 
 chaoticity parameter, $\kappa_0$, are specified in each figure.
 Analytical (wherever plotted) and numerical curves exactly overlap,
 and hence can not be seen separately.}
 \label{compk1tf4}
 \label{compk1tf1}
\end{figure}
It is straightforward to do time evolution now, for the state on the period-4 orbit, corresponding to the 
coherent state at $(0,0)$ which is $\otimes^3|0\kt$.
\begin{equation}
\label{eq29}
\begin{split}
&\mathcal{U}^{n} |000\rangle \equiv |\psi_n \kt =\frac{1}{2}e^{-i n \left(\frac{3 \pi}{4}+\kappa\right)}\left\lbrace (1+i^n) \left(
 \alpha_n |000\rangle  \right.  \right. \\ & \left. \left. + i \beta_n |\overline{W} \rangle
 \right) + (1-i^n) \left( i \alpha_n |111\rangle - \beta_n |W \rangle
 \right) \right\rbrace. 
 \end{split}
 \end{equation}
 From the $1$ and $2$ qubit reduced density matrices $\rho_1(n)=\text{tr}_{2,3} (|\psi_n \rangle \langle \psi_n |)$,
 $\rho_{12}(n)=\text{tr}_{3}( |\psi_n \rangle \langle \psi_n |)$, the entanglement of one qubit with the other
 two is measured by its entropy, and the entanglement between two qubits by the concurrence measure \cite{Wootters}.
 It turns out that for even values of the time $n$, $\rho_1(n)$ is diagonal and its eigenvalues are $\lambda(n,\kappa) = \frac{1}{2} U^{2}_{n-1}(\chi) $
 and $1-\lambda(n, \kappa)$, from which the linear entropy
 $S_{(0,0)}^{(3)}(n,\kappa)=2 \lambda(n,\kappa)(1- \lambda(n,\kappa))$ and its infinite time average of Eq.~(\ref{eq:avgpiby2})
follows. The two-qubit state is an ``X state" \cite{YuEberly2007} when the time is even, and results in the concurrence being \cite{SM}
    \begin{equation}
 \mathcal{C}(n)= \left| U_{n-1}(\chi) \right| \left| \frac{1}{2} 
 \vert U_{n-1}(\chi) \vert- \sqrt{1-\frac{3}{4} \vert U_{n-1}(\chi) \vert^2 } \right|.
\end{equation}

 Figure~\ref{compk1tf1} shows the comparison between these analytical results and those using experimental data from \cite{Neill16}, where two values
 of $\kappa_0$, $0.5$ and $2.5$ have been used.  The experimental data 
is the result of full state tomography and the procedure we have used to analyze
the data is outlined in the final section of the Supplementary materials \cite{SM}. The period-4 orbit is unstable at $\kappa_0=2.5$ and we see a rapid growth in the entanglement. However even at $\kappa_0=0.5$ entanglement grows to near maximal values, consistent with the large time average
 in Eq.~(\ref{eq:avgpiby2}) and Fig.~(\ref{fig:entropyplot}). We need use only even values of the time
 as for this state, $S(2n,\kappa)=S(2n-1, \kappa)$ and results in the steps of the top panel in Fig.~\ref{compk1tf1}. This is exact in the analytical expressions and quite remarkably present (but previously unnoticed) in the experimental data for the first few time steps. This curiosity results from $\mathcal{U}^{-1}|\psi_{2m}\kt$ being locally equivalent to $|\psi_{2m}\kt$. If $m$ itself is even, then it is straightforward to verify that applying the non-local part of the unitary operator $\mathcal{U}^{-1}$ results in $\otimes^3 e^{i \kappa \sigma_z}  |\psi_{2m} \rangle$, hence $|\psi_{2m-1}\kt$ is locally connected to $|\psi_{2m}\kt$, and all entanglement properties including concurrence
is left unchanged for an odd-to-even time step. A similar situation holds when $m$ is odd.

When the initial state is $\otimes^3|+\kt_y$ corresponding to the coherent state at $(\pi/2,\pi/2)$, the evolution lies entirely in the positive 
parity sector: $\mathcal{U}^n|+++\kt_y=$
\begin{equation}
 \frac{1}{2} e^{-i n \left(\frac{\pi}{4}+\kappa\right)}\left( (\alpha_n-i\sqrt{3}\beta_n^{*}) |\phi_1^{+} \rangle +  (\beta_n+i\sqrt{3}\alpha_n^{*})|\phi_2^{+} \rangle \right),
\label{eq:fixedptstate}
\end{equation}
Eigenvalues of the corresponding $\rho_1(n)$ are $\lambda(n, \kappa)=2\chi^2U_{n-1}(\chi)^2$ and $1-\lambda(n, \kappa)$, and the linear entropy is
\begin{equation}   
\label{eq-e2-8}
 S_{(\frac{\pi}{2}, \frac{\pi}{2})}^{(3)}(n,\kappa)= 4\chi^2U_{n-1}(\chi)^2 \left( 1-2\chi^2 U_{n-1}(\chi)^2 \right).
\end{equation}
See \cite{SM} for details. The plot showing comparison of linear entropy from experimental data and this expression for $\kappa_0=0.5$ 
and $\kappa_0=2.5$ are shown in Fig.~(\ref{compk1tf4}).
 It shows a much smaller growth for $\kappa=0.5$ in comparison 
to the state $|000\kt$, reflecting the stable neighborhood of $(\pi/2,\pi/2)$. This being consistent with the long time
average, already displayed in Eq.~(\ref{eq:avgpiby2}) which is derived from this expression. Qualitative discussions of the time-evolution have already been presented in \cite{Madhok2018_corr} and we move on to the 4-qubit case.

 \emph{Exact solution for four-qubits:} In this case the parity symmetry reduced and permutation symmetric basis
 in which $\mathcal{U}$ is block-diagonal is
 $\{ |\phi_1^{\pm} \rangle= \frac{1}{\sqrt{2}} (|W\rangle \mp | \overline{W} \rangle),\,
 |\phi_2^{\pm} \rangle = \frac{1}{\sqrt{2}} (|0000\rangle \pm | 1111 \rangle), \,  |\phi_3^{+} \rangle = \frac{1}{\sqrt{6}} \sum_{\mathcal{P}}|0011\rangle_{\mathcal{P}}\}$, where $|W\kt =\frac{1}{2}\sum_{\mathcal{P}}|0001\kt_{\mathcal{P}}$, $|\overline{W}\kt =\frac{1}{2}\sum_{\mathcal{P}}|1110\kt_{\mathcal{P}}$, and $\sum_{\mathcal{P}}$ sums over all possible permutations. A peculiarity of 4-qubits is that $|\phi_1^{+}\kt$ is an eigenstate of $\mathcal{U}$ with eigenvalue $-1$ for {\it all} values of the parameter $\kappa_0$. Thus the $5-$ dimensional space splits into $1\oplus2\oplus2$ subspaces on which the operators are $\mathcal{U}_0=-1$ and $\mathcal{U}_{\pm}$. 
  
The explicit form of powers of the $2 \times 2$ blocks of $\mathcal{U}$ are \cite{SM}
\begin{equation}
  \mathcal{U}_{+}^n
  =  e^{-\frac{i n}{2}(\pi+\kappa) }
  \begin{pmatrix}
   \alpha_n & i\beta_n^{*} \\  i\beta_n & \alpha_n^{*}
  \end{pmatrix}, \;\; \mbox{and}
\end{equation}
\begin{equation}
\label{eq12}
 \mathcal{U}_{-}^n = e^{-\frac{ 3i}{4}n \kappa} \left(
\begin{array}{cc}
 \cos \frac{n\pi}{2} & e^{\frac{3 i}{4}\kappa} \sin \frac{n\pi}{2} \\
 -e^{-\frac{3 i }{4} \kappa} \sin \frac{n\pi}{2}  &  \cos \frac{n\pi}{2} \\
\end{array}
\right)
\end{equation}
where $\alpha_n =  T_{n}(\chi)+\frac{i}{2}U_{n-1}(\chi)\cos{\kappa}$,
$\beta_n =  \frac{\sqrt{3}}{2}U_{n-1}(\chi)e^{i\kappa}$ and $\chi=\frac{1}{2}\sin{\kappa}$, with $\kappa=\kappa_0/2$. 
Using these it is possible to find the exact evolution of the entanglement entropy of any one-qubit and again in particular for the states $|0000\kt$, $|\pm\pm\pm\pm \kt_y$ this gives their long-time averaged linear entropy (for $\kappa_0 \neq 0, 2 \pi$) as \cite{SM}
\begin{equation*}
\br S^{(4)}_{(0,0)}(\kappa_0)\kt =\frac{1}{8}\left( \frac{9+2s_0}{3+s_0}\right),\, \br S^{(4)}_{(\frac{\pi}{2},\pm \frac{\pi}{2})}(\kappa_0)\kt=\frac{1}{8}\left(\frac{9-s_0}{3+s_0}\right).
\end{equation*}
where $s_0=\cos^2(\kappa_0/2)$.  Both reach their maximum value of $3/8$ when $\kappa_0=\pi$ and, remarkably, this matches with the average from the ensemble of random permutation symmetric states \cite{2018arXiv180600113S} of 4-qubits $S_{RMT}(4)$ as in the case of the 3-qubit case. In addition we  see that the average for the states at $(\pi/2, \pm \pi/2)$ attain the value of $1/4$ for arbitrarily small $\kappa_0$ in contrast to the 3-qubit case which vanishes as in Eq.~(\ref{eq:avgpiby2}). In fact the non-zero average is seen in numerical calculations to be attained only on averaging over extremely long times for small $\kappa_0$. 

\begin{figure}[h]
 \includegraphics{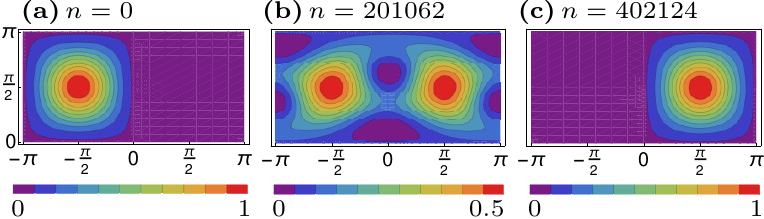}
 \caption{Husimi (quasi probability distribution) plots for the four-qubit initial state,  $\otimes^4|+\kt$, evolving under $n$ implementations of 
 $\mathcal{U}$, and leading to tunneling to the state,  $\otimes^4|-\kt$, at time $n_* \approx 128 \pi/\kappa_0^3 \approx 402124$. ($\kappa_0=0.1$). }
\label{fig:tunnel}
\end{figure}

This very slow process is due to tunneling between $\otimes^4|+\kt_y$ and $\otimes^4|-\kt_y$. At $\kappa_0=0$, two positive parity eigenvectors of $\mathcal{U}$,  $|\phi_1^+\kt$ and $|\phi_{23}^+\kt=\frac{1}{2} |\phi_2^{+}\rangle - \frac{\sqrt{3}}{2}|\phi_3^{+}\rangle$ are degenerate with eigenvalue $-1$. 
These can also be written as 4-qubit GHZ states \cite{GHZ0,GHZ}: $i |\phi_1^+\kt=\left(\otimes^4|+\kt_y -\otimes^4|-\kt_y \right)/\sqrt{2}$, the unchanging eigenstate,  and $|\phi_{23}\kt = \left(\otimes^4|+\kt_y +\otimes^4|-\kt_y \right)/\sqrt{2}$.
Thus
\begin{equation}
\mathcal{U}^n\otimes^4 |+\kt _y = (-1)^{n}\frac{i}{\sqrt{2}}|\phi_1^+\kt + \mathcal{U}_+^n \frac{1}{\sqrt{2}}|\phi_{23}^+\kt.
\label{eq:tunnelevolve}
\end{equation}
The eigenvalue of $\mathcal{U}_+$ that is $-1$ at $\kappa_0=0$ is $e^{i \gamma_{-}}$ with 
\begin{equation}
\gamma_{-}= \frac{\kappa_0}{4}+\pi -\sin^{-1}\left(\frac{1}{2}\sin \frac{\kappa_0}{2} \right) \approx \pi -\frac{\kappa_0^3}{128}.
\end{equation}
This implies that for $\kappa_0 \ll 1$, the corresponding state and $|\phi_1^+\kt$ are nearly degenerate. The splitting leads to a change in the relative phase of their contributions in Eq.~(\ref{eq:tunnelevolve})
and at time $n_* \approx 128 \pi/\kappa_0^3$ the evolved state is close to $\otimes^4|-\kt$, leading to tunneling as shown in Fig.~(\ref{fig:tunnel}) between what in the classical limit are two stable islands. At time $n=n_*/2$ the state obtained is close to the GHZ state $(\otimes^4 |+\kt_y -i \otimes^4 |-\kt_y)/\sqrt{2}$. 

This tunneling is observed whenever $\otimes^{2j}|\pm\kt$ are degenerate eigenstates of the rotation part of the Floquet $\mathcal{U}$. This implies that the number of qubits should be an integer multiple of $2\pi/p$, where
$p$ is the rotation angle (we have used $p=\pi/2$, and hence the tunneling occurs when the number of qubits is a multiple of 4).

\begin{figure}[h]
 \includegraphics{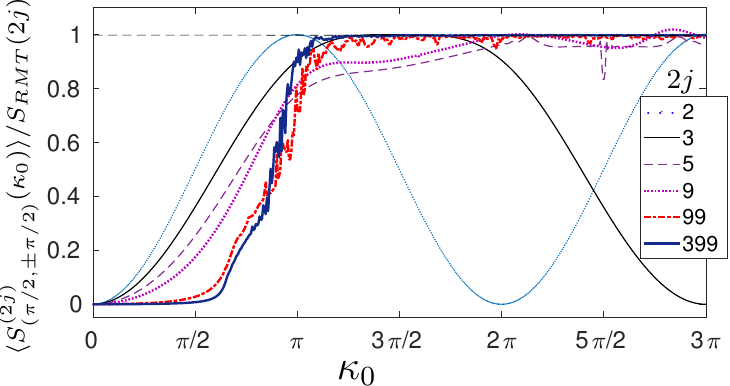}
 \caption{Normalized average single-qubit entanglement when the initial state is $\otimes^{2j}|+\kt_y$ for increasing number of qubits
 (except multiples of $4$ where there is tunneling for $p = \pi/2$.)}
\label{fig:avgentlargerj}
\end{figure}
For larger number of qubits, the average single-qubit entropy, normalized by the random state average,
is numerically found when the initial state is $\otimes^{2j}|+\kt_y$ and shown in Fig.~(\ref{fig:avgentlargerj}).
The trend is in keeping with a more complex classical phase space that becomes fully chaotic when the random state average is approached. The initial state being centered on a fixed point, increasing the number of qubits leads to a sharp growth beyond $\kappa_0=2$ when the fixed point becomes unstable, a more detailed  study of this is found in \cite{Bhosale-pre-2017}. Interestingly even for the 3-qubit case, for which we have the analytical evaluation in Eq.~(\ref{eq:avgpiby2}), a similar but smoother trend is displayed and reaches the random state value.

In summary, systems of few degrees of freedom, especially the exactly solvable 3- and 4- qubit instances of the kicked top provide insights
into how entropy and entanglement thermalize in closed quantum systems in the sense of long time averages approaching ensemble
averages. Experiments have already probed the 3-qubit case, which should be viewed as an integrable system. The 4-qubit case displays for the same rotation angle, tunneling and creation of GHZ states.   
Larger number of qubits can show genuine signatures of nonintegrability and chaos, and tunneling leads to creation of macroscopic superpositions that are generalized GHZ states. We hope our work raises new questions and adds to the discussion on the connections between integrability, quantum chaos, and thermalization. 

\begin{acknowledgments} We are grateful to the authors of \cite{Neill16} for generously sharing their experimental data, in particular to Pedram Roushan and Charles Neill for useful correspondence regarding the same.
\end{acknowledgments}  

\newpage
\section{Supplementary Material}
In this section, we provide details of the results obtained in the main text of the manuscript and give further evidence of thermalisation.
Last part of this supplemental material focuses on the 
analysis of the experimental data to obtain three-qubit 
density operators for various different initial states, 
after repeated implementations of the unitary operator $\mathcal{U}$.
\section{Three-qubit system under a kicked top Hamiltonian}
\begin{figure}
 \centering
 \includegraphics[scale=1,keepaspectratio=true]{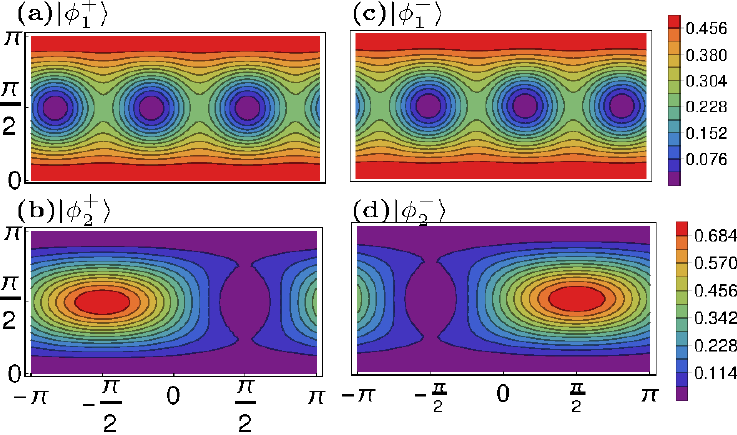}
 \caption{Husimi (quasiprobability distribution, $|\langle \phi_i|\theta_0,\phi_0 \rangle|^2$) plots
 for a set of four three-qubit bases states ($|\phi_i\rangle$), where $|\theta_0,\phi_0\rangle$ is 
 an arbitrary three-qubit, parametrized by ($\theta_0,\phi_0$).  \label{husimi3q}}
\end{figure}

This section contains a detailed description of the analytical 
solutions discussed in the main text file.
As per Eq.~(1) of the main text, unitary operator acting on a system of $3$-qubits,
that simulate the dynamics of a spin-$3/2$ under a kicked top Hamiltonian is given by,
\begin{eqnarray}
 \label{eq1a}
 \mathcal{U} = \exp && \left({-i \frac{\kappa_0}{6} (\sigma_1^z\sigma_2^z+\sigma_2^z\sigma_3^z+\sigma_3^z\sigma_1^z)} \right).
 \nonumber \\
 && \exp \left({-i \frac{\pi}{4}(\sigma_1^y+\sigma_2^y+\sigma_3^y)} \right), 
\end{eqnarray}
where $\kappa_0$ is the chaoticity parameter and 
$\sigma_l^{x,y,z}$ are the standard Pauli matrices.
Since $ \left[ \mathcal{U}, \sigma^y \otimes \sigma^y \otimes \sigma^y \right]=0$,
we obtain eigenvectors of $\sigma_{123}^y=\sigma^y \otimes \sigma^y \otimes \sigma^y$, that
block diagonalize $\mathcal{U}$.
Eigenvectors of $\sigma_{123}^y$ with eigenvalues $\pm1$ are given by
\begin{eqnarray}
\label{eq4}
 |\phi_1^{\pm} \rangle &=& \frac{1}{\sqrt{2}} (|000\rangle \mp i | 111 \rangle) \, \textrm{and} \nonumber  \\
  |\phi_2^{\pm} \rangle &=& \frac{1}{\sqrt{2}} (|W\rangle \pm i | \overline{W} \rangle), 
  \end{eqnarray}
where $|\overline{W}\rangle = \frac{1}{\sqrt{3}}(|011\rangle + |101\rangle + |110\rangle)$.
 Husimi plots for each of these bases vectors is shown in Fig.\ref{husimi3q}.
In this bases, the unitary operator `$\mathcal{U}$' is written as
\begin{equation} 
\label{eq6}
 \mathcal{U} = \begin{pmatrix}
            \mathcal{U}_{+} & 0_{2\times 2} \\ 0_{2\times 2} & \mathcal{U}_{-}
            \end{pmatrix}, 
\end{equation}
where $0_{2\times 2}$ is a null matrix,
and $2\times2$-dimensional block $\mathcal{U}_{+}$($\mathcal{U}_{-}$) is written 
in the bases $\{ \phi_{1}^{+}, \phi_{2}^{+} \}$ ($\{ \phi_{1}^{-}, \phi_{2}^{-} \}$),
thus being referred to as positive(negative)-parity subspace in our 
discussion. We have,
\begin{equation}
\label{eq7}
 \mathcal{U}_{\pm} = \begin{pmatrix}
            \langle \phi_{1}^{\pm} |\mathcal{U}| \phi_{1}^{\pm} \rangle & \langle \phi_{1}^{\pm} |\mathcal{U}| \phi_{2}^{\pm} \rangle \\
             \langle \phi_{2}^{\pm} |\mathcal{U}| \phi_{1}^{\pm} \rangle & \langle \phi_{2}^{\pm} |\mathcal{U}| \phi_{2}^{\pm} \rangle 
            \end{pmatrix}.
\end{equation}
This block diagonalization makes it easy to take the $n^{th}$ power of the
unitary operator $\mathcal{U}$,
\begin{equation}
\label{eq8a}
 \mathcal{U}^n = \begin{pmatrix}
            \mathcal{U}_{+}^n & 0_{2\times 2} \\ 0_{2\times 2} & \mathcal{U}_{-}^n
            \end{pmatrix}.
\end{equation}
The block operators $\mathcal{U}_{\pm}$ are explicitly 
found by using Eqs.(\ref{eq1a},~\ref{eq4},~\ref{eq7}). We have,
\begin{equation}
  \label{eq18a}
  \mathcal{U}_{\pm} = \pm  e^{\mp \frac{i \pi}{4}} e^{-i \kappa} \begin{pmatrix}
   \frac{i}{2}e^{-2i \kappa} & \mp \frac{\sqrt{3} }{2} e^{-2i \kappa} \\
   \pm \frac{\sqrt{3}}{2} e^{2i \kappa} &  -\frac{i}{2}e^{2i \kappa}
   \end{pmatrix},
 \end{equation}
 For simplicity the parameter $\kappa=\kappa_0/6$ is used in these expressions. 
 One can easily flip between $\mathcal{U}_{+}$ and $\mathcal{U}_{-}$ using
\begin{equation}
\label{eq13a}
 \mathcal{U}_{-}(\kappa)=\mathcal{U}_{+}^{*}(-\kappa),
\end{equation}
where $*$ is the conjugation operation \textit{in the standard bases}.
Re-writing $\mathcal{U^{\pm}}$ as a rotation by angle `$\theta$' about an arbitrary axis 
($\hat{\eta}=\sin{\alpha} \cos{\beta} \hat{x}
 + \sin{\alpha} \sin{\beta} \hat{y} + \cos{\alpha} \hat{z}$),
\begin{equation}
\label{eq15}
 \mathcal{U}_{+}\dot{=} e^{-i \theta \sigma^{\hat{\eta}}} = 
 \exp[-i \theta (\sin{\alpha} \cos{\beta} \sigma^x
 + \sin{\alpha} \sin{\beta} \sigma^y + \cos{\alpha} \sigma^z)],
\end{equation}
which is valid upto phases.
On comparison with Eq.~\ref{eq18a}, we obtain,
$ \cos{\theta} =\frac{1}{2} \sin{2\kappa}$, 
 $\beta=\pi/2 +2 \kappa$, and 
 $\sin{\alpha} = \sqrt{3}/(2\sin{\theta})$.
 To evolve initial states we need $\mathcal{U}^n$ and therefore
$\mathcal{U}_{\pm}^n$,
  \begin{eqnarray}
 \label{eq23}
  \mathcal{U}_{+}^n &=& e^{-i \frac{n\pi}{4}}e^{-i n \kappa}. \nonumber \\
  && \begin{pmatrix}
   \cos{n\theta}- i \sin{n\theta} \cos{\alpha} & -i \sin{n\theta} \sin{\alpha}\,e^{-i \beta} \\
   -i \sin{n\theta} \sin{\alpha}\,e^{i \beta} &  \cos{n\theta}+ i \sin{n\theta} \cos{\alpha}
  \end{pmatrix}. \nonumber \\
   \end{eqnarray}
   Further, $\cos{(n\theta)}$ and $\sin{(n\theta)}/\sin{\theta}$ are identified
   as the Chebyshev polynomials of first kind ($T_n(\chi)$) and 
   second kind ($U_{n-1}({\chi})$) respectively with 
   $\chi=\cos{\theta}=\sin(2\kappa)/2$. Re-writing $\mathcal{U}_{\pm}^n$ in 
   a more convenient form,
\begin{equation}
\label{eq:Upluspowern}
\mathcal{U}_{\pm}^n = (\pm 1)^n e^{-i n (\pm  \frac{\pi}{4}+\kappa)}
\begin{pmatrix}
 \alpha_n &
   \mp \beta_n^* \\
   \pm \beta_n &  
   \alpha_n^*
 \end{pmatrix}, 
\end{equation}   
   where $\alpha_n = T_n(\chi)+\frac{i}{2}\, U_{n-1}(\chi) \cos 2\kappa $ and 
 $ \beta_{n} = (\sqrt{3}/2)\, U_{n-1}(\chi) \,e^{2i \kappa}$. 
\subsection{Special case 1: $|\psi_0 \rangle=|000\rangle$}
Considering a three-qubit state $|\psi_0 \rangle=|000\rangle$,
under the action of $n$ implementations of $U$,
\begin{eqnarray}
\label{eq9}
 \mathcal{U}^n |000\rangle &=& \frac{1}{\sqrt{2}} \mathcal{U}^n \left( |\phi_{1}^{+} \rangle + |\phi_{1}^{-} \rangle \right) \nonumber  \\
                 &=& \frac{1}{\sqrt{2}} \left( \mathcal{U}_{+}^n |\phi_{1}^{+} \rangle + \mathcal{U}_{-}^n |\phi_{1}^{-} \rangle \right).
\end{eqnarray}

Final state of three-qubit system ($|\psi_n\rangle$), after $n$ implementations of 
the unitary operator $\mathcal{U}$ on $|\psi_0\rangle$, is
given by (from Eq.(\ref{eq:Upluspowern})),
\begin{equation}
\label{eq29}
\begin{split}
&\mathcal{U}^{n} |000\rangle \equiv |\psi_n \kt =\frac{1}{2}e^{-i n \left(\frac{3 \pi}{4}+\kappa\right)}\left\lbrace (1+i^n) \left(
 \alpha_n |000\rangle  \right.  \right. \\ & \left. \left. + i \beta_n |\overline{W} \rangle
 \right) + (1-i^n) \left( i \alpha_n |111\rangle - \beta_n |W \rangle
 \right) \right\rbrace. 
 \end{split}
 \end{equation}
We further study correlations such as linear entropy of the 
single-qubit reduced state ($\rho_1$) of 
$\rho_n=|\psi_n\rangle \langle \psi_n |$
and concurrence ($\mathcal{C}$) between any two qubits, 
\subsubsection{Linear entropy (for even values of `$n$') \label{sec1a}}
Single-party reduced density operator
is obtained by tracing out any of the two qubits of 
the three-qubit density operator, 
\[\rho_1=\textrm{Tr}_{2,3} |\psi_n \rangle \langle \psi_n |\]
For even values of `$n$', eigenvalues of the single qubit state are, 
$\lambda_n^{(1)}=\lambda_n$ and $\lambda_n^{(2)}=1-\lambda_{n}$, where $\lambda_n=\frac{2}{3}|\beta_n|^2$.
Therefore the measure of entanglement based on linear entropy,
\begin{equation}
  S^{(3)}_{(0,0)}(n,\kappa)=1- \sum_{i=1}^2 \left(\lambda_n^{(i)}\right)^2 = 2 (\lambda_n^{})(1- \lambda_n^{}),
 \label{ent1}
\end{equation}

where, $\lambda_n = \frac{1}{2} U^{2}_{n-1}(\chi) $.
It is also interesting to look at long-time averaged linear entropy.
Re-writing Eq.~\ref{ent1} as,
\begin{eqnarray} \label{ent4}
 S^{(3)}_{(0,0)}(n,\kappa) &=& U^{2}_{n-1}(\chi)-\frac{1}{2} U^{4}_{n-1}(\chi) \\
 &=& \frac{\sin^2n\theta}{\sin^2\theta} -\frac{1}{2} \frac{\sin^4n\theta}{\sin^4\theta},
\end{eqnarray}
Long-time averaged linear entropy,
\begin{eqnarray} \label{ent5}
  \br S^{(3)}_{(0,0)}(\kappa) \kt &=& \frac{1}{\sin^2\theta} \br \sin^2n\theta \kt -\frac{1}{2\sin^4\theta} \br \sin^4n\theta \kt, \\
 &=& \frac{1}{2\sin^2\theta}  -\frac{3}{16\sin^4\theta}.
\end{eqnarray}
Further, using $\cos \theta = \frac{1}{2}\sin 2\kappa$, we obtain,
  \begin{equation}  \label{ent6}
  \br S^{(3)}_{(0,0)}(\kappa) \kt=\frac{5-2\sin^2(2\kappa)}{\left(4-\sin^2(2\kappa)\right)^2},
 \end{equation}
that attains its maximum value of $1/3$ at $\kappa=\pi/4$.

\subsubsection{Concurrence (for even values of `$n$')}
\begin{equation}
 \rho_{12}= \begin{pmatrix}
 |\alpha_n|^2 & 0 & 0 & -\frac{i}{\sqrt{3}} \alpha_n \beta_n^{*} \\
 0 & \frac{1}{3} |\beta_n|^2 & \frac{1}{3} |\beta_n|^2 & 0 \\
  0 & \frac{1}{3} |\beta_n|^2 & \frac{1}{3} |\beta_n|^2 & 0 \\
\frac{i}{\sqrt{3}} \alpha_n^{*} \beta_n & 0 & 0 & \frac{1}{3}|\beta_n|^2
 \end{pmatrix},
\end{equation}
which is an `$X$' state, whose concurrence, $\mathcal{C}(\rho_{12})$ is measured by \cite{YuEberly2007},
$2.  max \left[ 0, \frac{1}{3}|\beta_{n}|^{2} - \frac{1}{\sqrt{3}} |\alpha_{n}||\beta_{n}|,
  -( \frac{1}{3}|\beta|_{n}^{2} - \frac{1}{\sqrt{3}} |\alpha_{n}||\beta_{n}|)\right]$. Thus,
  \begin{equation}
    \mathcal{C}(\rho_{12})  =2 \left| \frac{1}{3} \vert \beta_n \vert^2
 -\frac{1}{\sqrt{3}} \vert \alpha_n \vert \vert \beta_n \vert \right|. 
  \end{equation}
Substituting the values of $\alpha_n$ and $\beta_n$, concurrence is
given by
    \begin{equation}
 \mathcal{C}(\rho_{12})= \left| U_{n-1}(\chi) \right| \left| \frac{1}{2} 
 \vert U_{n-1}(\chi) \vert- \sqrt{1-\frac{3}{4} \vert U_{n-1}(\chi) \vert^2 } \right|.
\end{equation}
\subsubsection{Correlations for odd values of `$n$'}

To obtain the values of linear entropy and concurrence for 
states (in Eq.(\ref{eq29})) for odd values of $n$,
one can evolve the even $n=2m$ states one step backward or forward
in time, such as,
\begin{equation}
 |\phi_{2m-1}\rangle=\mathcal{U}^{-1} |\phi_{2m}\rangle,
\end{equation}
where $\mathcal{U}$ is the unitary operator (given in Eq.\ref{eq1a}).
Considering the backward evolution of $|\phi_{2m}\rangle$
(say for even value of $m$),
under the non-local part of the unitary operator $\mathcal{U}$,
\begin{eqnarray}
  |\psi_{2m-1}\rangle &=&  
  e^{i \kappa (\sigma_1^z\sigma_2^z+\sigma_2^z\sigma_3^z+\sigma_3^z\sigma_1^z)} 
  \left(
 \alpha_{2m} |000\rangle + i \beta_{2m} |\overline{W} \rangle
 \right), \nonumber \\
  &=& e^{3i \kappa} \alpha_{2m} |000\rangle + i e^{-i \kappa} \beta_{2m} |\overline{W} \rangle, \nonumber \\
 &=&  \mathcal{V}\otimes \mathcal{V}\otimes \mathcal{V} |\phi_{2m} \rangle,
\end{eqnarray}
where single qubit unitary operator $\mathcal{V} = e^{i \kappa \sigma_z}$.
Thus the three qubit state $|\psi_0\rangle$, after odd 
numbered implementations of the unitary operator $\mathcal{U}$ are 
locally connected to the state obtained after even numbered
implementations of the operator $\mathcal{U}$. 
\subsection{Special case 2: $|\psi_0 \rangle=|+++\rangle$}
Considering a three-qubit state, $|\psi_0\rangle=|+++\rangle$, where
$|+\rangle=\frac{1}{\sqrt{2}}(|0\rangle+i|1\rangle)$
is an eigenvector of $\sigma_y$ with eigenvalue $+1$.
Three qubit state is explicitly written as
\begin{equation}
 |\psi_0\rangle = \frac{1}{2} \left( |\phi_1^{+} \rangle + i \sqrt{3} |\phi_2^{+} \rangle \right), \label{b1}
\end{equation}
which lies in the positive parity subspace.
Three qubit state after $n$ implementations of $\mathcal{U}$ is given by
(upto an overall phase),
\begin{equation}
 |\psi_n\rangle = \frac{1}{2} e^{-in(\frac{\pi}{4}+\kappa)}\left( \gamma_n |\phi_1^{+} \rangle + \delta_n |\phi_2^{+} \rangle \right),
 \label{eq-e2-4}
\end{equation}
where $\gamma_n=\alpha_n-i\sqrt{3}\beta_n^{*}$ and $\delta_n=\beta_n+i\sqrt{3}\alpha_n^{*}$.
One can obtain the single-party reduced state by tracing out any two-qubits,
\begin{eqnarray}
&& \rho_{A} = \frac{1}{4}. \nonumber \\
&& \begin{pmatrix}
           \frac{1}{2} \left( |\gamma_n|^2 + |\delta_n|^2 \right) 
           & -\frac{i}{3} \left( |\delta_n|^2 - \sqrt{3}\; Im( \gamma_n \delta_n^{*} ) \right) \\
           \frac{i}{3} \left( |\delta_n|^2 - \sqrt{3}\; Im( \gamma_n \delta_n^{*} ) \right) &
           \frac{1}{2} \left( |\gamma_n|^2 + |\delta_n|^2 \right) 
            \end{pmatrix}.  \nonumber \\ \label{eq-e2-5}
\end{eqnarray}
Eigenvalues of $\rho_A$ are $\frac{1}{2} \pm  |\rho_A(1,2)|$,
which are explicitly given by
\begin{equation}
\lambda_n^{(1)}=2\chi^2U_{n-1}(\chi)^2 \, \rm{and} \, \lambda_n^{(2)}= 1-2\chi^2U_{n-1}(\chi)^2.
 \label{eq-e2-6}
\end{equation}
Linear entropy of this single-party reduced state is found to be
\begin{equation}
 S^{(3)}_{(\frac{\pi}{2},-\frac{\pi}{2})}(n,\kappa)=\frac{1}{2}-2 |\rho_A(1,2)|^2=2\lambda_n(1-\lambda_n),
  \label{eq-e2-7}
\end{equation}
which in a much simplified form is given by
\begin{equation}   \label{eq-e2-8}
S^{(3)}_{(\frac{\pi}{2},-\frac{\pi}{2})}(n,\kappa)= 4\chi^2U_{n-1}(\chi)^2 \left( 1-2\chi^2U_{n-1}(\chi)^2 \right).
\end{equation}
We also obtain the 
long time-average value of the linear entropy, given by,
 \begin{equation}
  \br S^{(3)}_{(\frac{\pi}{2},-\frac{\pi}{2})}(\kappa) \kt=\frac{\sin^2(2\kappa)}{\left(4-\sin^2(2\kappa)\right)^2}\left( 8-5\sin^2(2\kappa)\right),
 \end{equation}
which, when $\kappa=\pi/4$ approaches $1/3$. 
Coincidentaly, this is same as the average linear entropy 
of a single-qubit reduced state in a set of random symmetric
three-qubit states. 
\subsection{Linear entropy of an arbitrary three-qubit 
permutation symmetric state}
\begin{figure}
\includegraphics[scale=0.9]{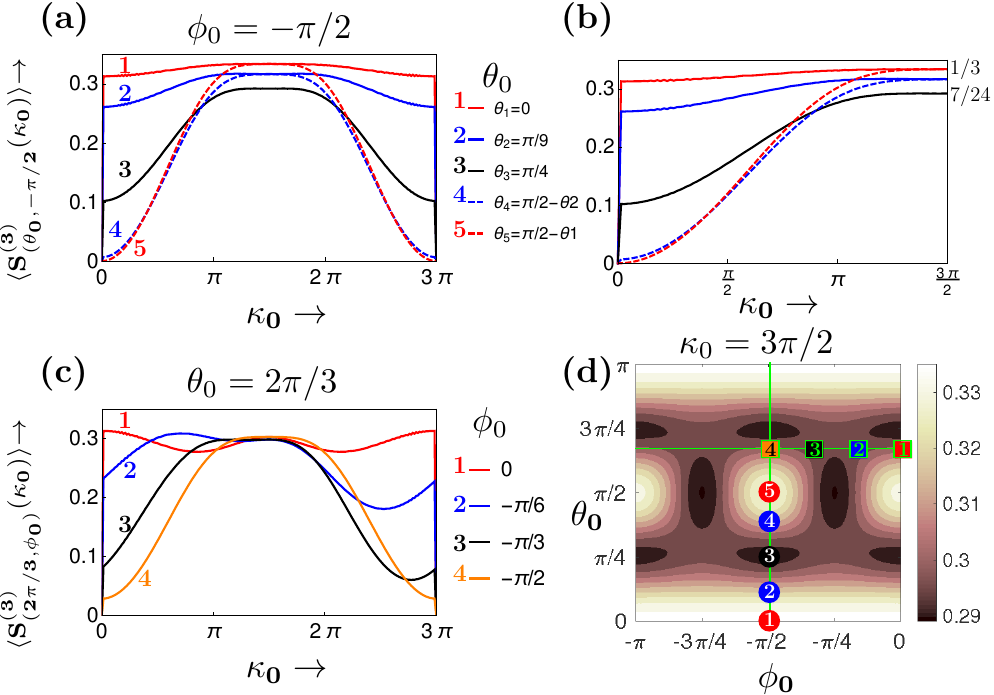}
\caption{(a),(b) Time averaged linear entropy ($\langle  S_{(\theta_0,-\pi/2)}^{(3)} \rangle$)
 of a single party reduced state vs chaoticity parameter $\kappa_0$. 
 Different curves correspond to different
 initial states, $|\theta_0,-\pi/2\rangle$ as labelled $1$ to $5$, 
 alongwith explicit values of $\theta_0$ given in the plot legends.
 These corresponding initial states $|\theta_0,\phi_0\kt$ are also 
 marked as numbered circles in the contour plot given in part (d).
 Part (c) contains the plots for $\langle  S_{(2\pi/3,\phi_0)}^{(3)} \rangle$
 vs chaoticity parameter $\kappa_0$ for a fixed value of $\theta_0=2\pi/3$. 
 Different curves correspond to different
 initial states, labelled by numbers $1$ to $4$
 alongwith explicit values of $\phi_0$ given in the plot legends.
 Respective initial states $|\theta_0,\phi_0\kt$ are also 
 marked as numbered squares (with a green border) in the contour
 plot given in part (d). Contour plot shown in part (d) corresponds to 
 $\kappa_0=3\pi/2$. \label{entropyplot}}
\end{figure}
Considering a three-qubit state
\begin{equation}
\label{eqg1n}
 |\psi_0\rangle= 
 a_1 |\phi_1^{+}\rangle + a_2 |\phi_2^{+}\rangle +
 b_1 |\phi_1^{-}\rangle + b_2 |\phi_2^{-}\rangle.
\end{equation}
Each of the three qubits are initialized in the same state 
($|\psi \kt=\cos \frac{\theta_0}{2} |0\kt + e^{-i \phi_0} \sin \frac{\theta_0}{2} |1\kt$, 
in the computational bases), such that the initial state of the
$3-$ qubit system is $|\psi_0 \kt=\otimes^{3} |\psi \kt$,
where $\theta_0 \in [0,\pi]$ and $\phi_0 \in [-\pi, \pi]$.
Repeated implementations of the unitary operator $\mathcal{U}$, 
leads to $|\psi_n\kt=\mathcal{U}^n|\psi_0\kt$.
We obtain single-party reduced density operator
 by tracing out any of the two qubits of 
the three-qubit density operator ($\rho_n=|\psi_n\kt \langle \psi_n|$),
leading to,
\begin{equation} \label{eqa1}
 \rho_i = \begin{pmatrix}
             r & s \\ s^{*} & 1-r 
             \end{pmatrix},
\end{equation}
where the elements of the density operator are
given by
\begin{eqnarray}
 r &=& \frac{1}{2}+\textrm{Re}\left( a_{1n}b_{1n}^{*} + \frac{1}{3}a_{2n}b_{2n}^{*} \right)  
 \quad \textrm{and}\nonumber \\ 
 s &=& \frac{1}{\sqrt{3}} \textrm{Re}\left(a_{1n}b_{2n}^{*} + b_{1n}a_{2n}^{*} \right) 
 + \frac{i}{\sqrt{3}} \textrm{Im}\left( a_{1n}a_{2n}^{*} + b_{1n}b_{2n}^{*} \right) \nonumber \\ 
 & & -\frac{i}{3} \left( a_{2n} + b_{2n} \right) \left( a_{2n}^{*} - b_{2n}^{*} \right).
\end{eqnarray}
Where the coefficients, $a_{1n}=a_1  \alpha_n -  a_2 \beta_n^{*}$, 
$a_{2n}=a_1 \beta_n +  a_2 \alpha_n^{*}$, $b_{1n}=i^{n} \left(
 b_1  \alpha_n + b_2 \beta_n^{*} \right)$, and $b_{2n}=i^{n} \left(
 b_2  \alpha_n^{*} - b_1 \beta_n \right)$.
Linear entropy of the single-qubit (Eq.(\ref{eqa1})) is thus given by,
\begin{equation} \label{eqa2}
 S_{(\theta_0,\phi_0)}^{(3)}(n,\kappa)=2 \left[ r(1-r)-|s|^2 \right].
\end{equation}
Thus linear entropy is obtained as a function of the initial-state parameters ($\theta_0, \phi_0$).
Long time average linear entropy is calculated numerically with $n=1000$
for various initial states as shown in Fig~\ref{entropyplot}.
Part (a) and (c) of Fig.~\ref{entropyplot} show the variation of 
time average entropy with chaoticity parameter for a period $2\pi j$.
Pairs of complimentary $\theta_0$s, saturate to same values 
in the region around $\kappa_0=3\pi/2$. 
Part (b) of Fig~\ref{entropyplot} highlights the range of values of 
average linear entropy at $\kappa_0=3\pi/2$, a scale of similar range 
in part (d) depicts that the linear entropy of a single-qubit reduced 
state for an arbitrary value of parameters ($\theta_0, \phi_0$) fall 
into this range.
\par
Further, we have obtained an explicit closed form experssion for 
long time average linear entropy for an arbitrary ($\theta_0, \phi_0$)
at $\kappa_0=3\pi/2$, which is discussed in the main text.
\section{Four qubit kicked top}
\begin{figure}
 \centering
 \includegraphics[scale=1,keepaspectratio=true]{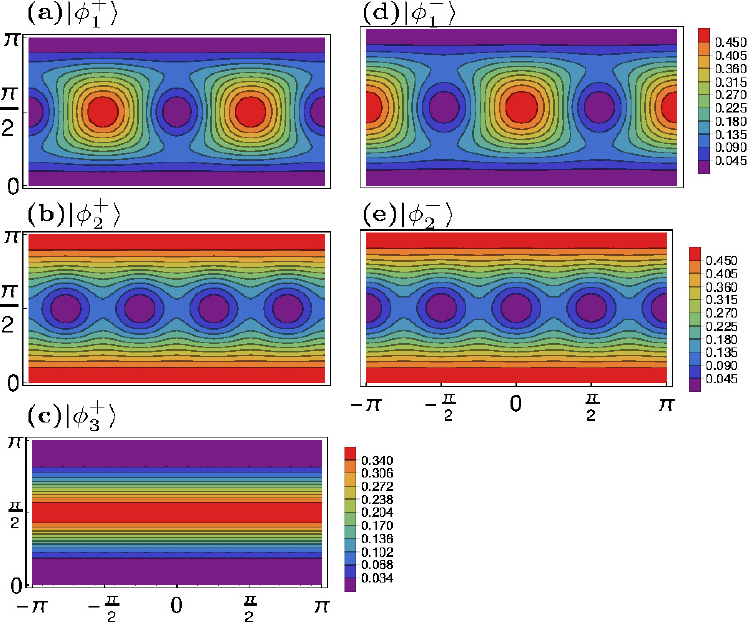}
 \caption{Husimi (quasiprobability distribution, $|\langle \phi_i|\theta_0,\phi_0 \rangle|^2$) plots
 for a set of five four-qubit bases states ($|\phi_i\rangle$), where $|\theta_0,\phi_0\rangle$ is 
 an arbitrary four-qubit, parametrized by ($\theta_0,\phi_0$).}
 \label{husimi4q}
\end{figure}

Considering a spin-$2$ system, whose dynamics is effectively
simulated by a four-qubit system, confined to its five-dimensional
symmetric subspace. 
Re-writing Eq.~(1) from the main text, explicitly for a 
system of four qubits,
\begin{eqnarray}
\label{eq1}
 \mathcal{U} &=& \exp \left(-{i \frac{\kappa_0}{8} (\sigma_1^z\sigma_2^z+\sigma_1^z\sigma_3^z
 +\sigma_1^z\sigma_4^z+\sigma_2^z\sigma_3^z+\sigma_2^z\sigma_4^z+\sigma_3^z\sigma_4^z)} \right) \nonumber
 \\ &&
 \exp \left({-i \frac{\pi}{4}(\sigma_1^y+\sigma_2^y+\sigma_3^y+\sigma_4^y)} \right),
\end{eqnarray}
where all the terms have their usual meanings.
We have,
\begin{equation}
\label{eq2}
 \left[  \mathcal{U}, \sigma^y \otimes \sigma^y \otimes \sigma^y \otimes \sigma^y \right]=0.
\end{equation}
Unitary operator `$\mathcal{U}$' becomes block diagonal in the eigenbases of 
operator $\sigma_{1234}^y=\otimes^{4} \sigma^y$, given by,
\begin{eqnarray}
\label{eq4}
 |\phi_1^{\pm} \rangle &=& \frac{1}{\sqrt{2}} (|W\rangle \mp | \overline{W} \rangle) \nonumber \\
 &=& \frac{1}{\sqrt{2}} \left( \frac{1}{2} \sum_{\mathcal{P}}|0001\rangle_{\mathcal{P}} \mp 
 \frac{1}{2} \sum_{ \mathcal{P}}|0111\rangle_{\mathcal{P}} \right), \nonumber  \\
 |\phi_2^{\pm} \rangle &=& \frac{1}{\sqrt{2}} (|0000\rangle \pm | 1111 \rangle), \, \textrm{and} \nonumber  \\
  |\phi_3^{+} \rangle &=&  \frac{1}{\sqrt{6}} \sum_{\mathcal{P}}|0011\rangle_{\mathcal{P}}, 
  \end{eqnarray}
  where $\sum_{\mathcal{P}}$ sums over all possible permutations.
  Eigenvectors of $\otimes^{4} \sigma^y$, $|\phi_i^{+}\rangle$ with eigenvalues $+1$,
  lie in the positive parity subspace, while $|\phi_i^{-}\rangle$ with eigenvalues $-1$
  belong to the negative-parity subspace.
  Husimi plots for each of these bases vectors is shown in Fig.\ref{husimi4q}.
  It is interesting to note that $|\phi_1^{+}\rangle$ is also an eigenvector of 
  $\mathcal{U_{+}}$ with eigenvalue $-1$.
 In this set of bases, the unitary operator `$\mathcal{U}$' becomes block diagonal, 
which makes it easy to take the $n^{th}$ power of the
unitary operator $ \mathcal{U}$,
\begin{equation}
\label{eq8}
  \mathcal{U}^n = \begin{pmatrix}
           \mathcal{U}_0^n & 0_{1\times 2}& 0_{1\times 2} \\ 0_{2\times 1} &   
           \mathcal{U}_{+}^n & 0_{2\times 2} \\ 0_{2\times 1} & 0_{2\times 2} &  \mathcal{U}_{-}^n
            \end{pmatrix}, 
\end{equation}
This simplifies our problem to much extent, which is now decomposed
to work only in the $2\times2$-dimensional subspaces.

\par
Various blocks are written here explicitly, we have
\begin{equation}
 \mathcal{U}_0=\langle \phi_1^{+} |\mathcal{U}|\phi_1^{+} \rangle = -1,
\end{equation}
which is a part of the positive-parity subspace.
Block $\mathcal{U}_{+}$ is written in the bases $\{\phi_2^{+},\phi_3^{+}\}$,
\begin{equation}
\label{eq12}
  \mathcal{U}_{+} = -ie^{-\frac{i \kappa }{2}} \left(
\begin{array}{cc}
 \frac{i}{2} e^{-i \kappa} & \frac{\sqrt{3}i}{2}  e^{-i \kappa} \\
 \frac{\sqrt{3}i}{2}  e^{i \kappa} & -\frac{i}{2} e^{i \kappa} \\
\end{array}
\right).
\end{equation}
Block $\mathcal{U}_{-}$ is written in the bases $\{\phi_1^{-},\phi_2^{-}\}$,
\begin{equation}
\label{eq12}
  \mathcal{U}_{-} = e^{-\frac{3 i \kappa }{4}} \left(
\begin{array}{cc}
 0 & e^{\frac{3 i \kappa }{4}} \\
 -e^{-\frac{3 i \kappa }{4}} & 0 \\
\end{array}
\right)
\end{equation}
Re-writing as a rotation ($e^{i \theta \sigma^{\hat{\eta}}}$) by angle `$\theta$' about an arbitrary axis 
($\hat{\eta}=\sin{\alpha} \cos{\beta} \hat{x}
 + \sin{\alpha} \sin{\beta} \hat{y} + \cos{\alpha} \hat{z}$).
 The generator of this rotation being, $\sigma^{\hat{\eta}}=\sin{\alpha} \cos{\beta} \sigma^{x}
 + \sin{\alpha} \sin{\beta} \sigma^{y} + \cos{\alpha} \sigma^{z}$. 
 A general rotation operator, raised to power `$n$' is thus of the form $e^{i n\theta \sigma^{\hat{\eta}}}$.
 We have, 
 \begin{eqnarray}
   \mathcal{U}_{+}^n &=& e^{-\frac{i n(\pi+\kappa) }{2}} e^{i n\theta \sigma^{\hat{\eta}}} \nonumber \\
  &=&  e^{-\frac{i n(\pi+\kappa) }{2}} 
  \begin{pmatrix}
   \cos{n\theta}+\frac{i}{2}\frac{\sin{n\theta}}{\sin{\theta}}\cos{\kappa} & 
   \frac{i\sqrt{3}}{2}\frac{\sin{n\theta}}{\sin{\theta}}e^{-i\kappa} \\
   \frac{i\sqrt{3}}{2}\frac{\sin{n\theta}}{\sin{\theta}}e^{i\kappa} & 
   \cos{n\theta}-\frac{i}{2}\frac{\sin{n\theta}}{\sin{\theta}}\cos{\kappa} \nonumber \\
  \end{pmatrix},
 \end{eqnarray}
where $\cos{\theta}=\sin{\kappa}/2$, $\beta=\kappa$, $\sin{\alpha}=\sqrt{3}/(2\sin{\theta})$,
and $\cos{\alpha}=\cos{\kappa}/(2\sin{\theta})$. Further simplification leads to the form,
\begin{eqnarray}
  \mathcal{U}_{+}^n &=&  e^{-\frac{i n(\pi+\kappa) }{2}} 
  \begin{pmatrix}
   \alpha_n & i\beta_n^{*} \\  i\beta_n & \alpha_n^{*}
  \end{pmatrix},
 \end{eqnarray}
 such that, 
 \begin{eqnarray}
\alpha_n &=&  T_{n}(\chi)+\frac{i}{2}U_{n-1}(\chi)\cos{\kappa}  \qquad \textrm{and} \nonumber \\
\beta_n  &=&  \frac{\sqrt{3}}{2}U_{n-1}(\chi)e^{i\kappa},
 \end{eqnarray}
 where $T_{n}(\chi)$ and $U_{n-1}(\chi)$ are the Chebyshev polynomials of first 
 and second kinds respectively, with $\chi=\sin{\kappa}/2$.
 
 \par
 Further, comparing $ \mathcal{U}_3$ with 
 the general rotation operator, we obtain, $\theta=\pi/2, \alpha=\pi/2, \beta=-(\frac{\pi}{2}+\frac{3\kappa}{4})$.
 Thus,
\begin{equation}
\label{eq12}
  \mathcal{U}_{-}^n = e^{-\frac{ 3in \kappa }{4}} \left(
\begin{array}{cc}
 \cos \frac{n\pi}{2} & e^{\frac{3 i \kappa}{4}} \sin \frac{n\pi}{2} \\
 -e^{-\frac{3 i \kappa }{4}} \sin \frac{n\pi}{2}  &  \cos \frac{n\pi}{2} \\
\end{array}
\right)
\end{equation}
\subsection{Special case 1: $|\psi_0 \rangle=|0000\rangle$}
Considering four qubit state $|0000\rangle$, under the `$n$'
implementations of unitary operator $\mathcal{U}$,
\[ \mathcal{U}^n |0000\rangle = \frac{1}{\sqrt{2}} \left( \mathcal{U}^n_{+} |\phi_2^{+}\rangle
+ \mathcal{U}^n_{-} |\phi_2^{-}\rangle \right), \]
leading to final state $|\psi_n\rangle$. We further analyse 
single-qubit reduced density operator ($\rho_1$) of the final state 
($\rho_n=|\psi_n\rangle \langle \psi_n|$) by tracing out 
any three qubits of the four qubit system. Single-qubit reduced 
density operator obtained in this case is diagonal for even values of $n$,
eigenvalues being $\lambda$ and $1-\lambda$, where 
$\lambda=\frac{1}{2}\left( 1+\textrm{Re} \left( \alpha_n e^{in\kappa_0/4} \right) \right)$.
For even values of $n$, linear entropy of a single-qubit 
reduced state is given by,
\begin{equation}
  S_{(0,0)}^{(4)}(n,\kappa_0)=\frac{1}{2}\left[ 1- \left( \textrm{Re} \left( \alpha_n e^{in\kappa_0/4} \right) \right)^2 \right].
\end{equation}
Long time average of the linear entropy is obtained by averaging over $n$
as shown in Section~\ref{sec1a}. Final expression for $\langle S_{(0,0)}^{(4)} \rangle$
thus obtained, is discussed in the main text.
\subsection{Special case 2: $|\psi_0 \rangle=|++++\rangle$}
This state lies entirely in the positive parity subspace of 
our five dimensional permutation symmetric space of four qubits,
given by
\begin{equation}
\otimes^4 |+\kt = \frac{i}{\sqrt{2}}|\phi^+_1\kt -\frac{1}{\sqrt{8}} |\phi_2^{+}\rangle + \sqrt{\frac{3}{8}}|\phi_3^{+}\rangle,
\end{equation}
which under the action of $\mathcal{U}^n$, leads to $|\psi_n\rangle=\mathcal{U_{+}}^n|++++\rangle$. 
Reduced density operator of each of these four qubits is given by,
\begin{equation}
 \rho_1=\textrm{Tr}_{2,3,4}\left( |\psi_n\rangle\langle \psi_n | \right) = 
 \begin{pmatrix}
  r_{(4)} & s_{(4)} \\ s_{(4)}^{*} & 1-r_{(4)}
 \end{pmatrix},
\end{equation}
where, diagonal element, $r_{(4)}=1/2$ and $s_{(4)}=\frac{i(-1)^n}{2} \left( \sin(\delta) U_{n-1}(\chi) 
\cos(\kappa_0/2) - T_n(\chi) \cos(\delta)\right)$ with $\delta=-n\left( 2\pi+\kappa_0\right)/4$.
A closed form expression for long time average linear entropy is then obtained using,
\[ S^{(4)}_{(\pi/2,-\pi/2)}= 2\left( \langle r_{(4)} \rangle - \langle r_{(4)}^2 \rangle  - \langle |s_{(4)}|^2 \rangle \right), \]
which is discussed in the main text.

\section{Experimental state reconstruction}
We analyse the data from a recent experiment~\cite{Neill16}, 
that demonstrates the kicked top dynamics of a spin-$3/2$,
using three superconducting transmon qubits.
State of a three-qubit system is obtained via complete quantum state 
tomography using a set of 64 projective measurements. These projective 
measurements are constructed by taking the combinations of Pauli-$x,y,z$ 
matrices ($\sigma_x$, $\sigma_y$ $\sigma_z$) and the Identity 
operator ($I$)~\cite{james-pra-2001, Neill16}.
These measurements are experimentally realized by various
single-qubit rotations ($\mathcal{R}$) followed by $\sigma_z$ measurements
on individual qubits, that effectively performs a $\sigma_i'$
measurement (for $i'=x$, $\mathcal{R}=$Hadamard operator $(Hd)$; 
$i'=y$, $\mathcal{R}=$ Phase shift $(\mathcal{S}).Hd$; $i'=z$, 
$\mathcal{R}=I$)~\cite{Neill16}.
Multiple implementations of each measurement, provides the  
relative occupancy of the eight bases states of a three-qubit system.
 The resulting relative populations ($p_m$) of the eight bases
 states are thus obtained experimentally. In order to 
compensate the effect of errors induced by the measurements,
 the intrinsic populations ($p_{int}$) are obtained
 via a correction matrix ($F$)~\cite{steffen-science-2006, lucero-prl-2008}. 
 We have, $p_{int}=F^{-1} p_m$, where $F=F_1 \otimes F_2 \otimes F_3$.
 $F_i$ is the measurement error corresponding to $i^{th}$ qubit,
 given as,
 \begin{displaymath}
  F_i= \left( \begin{array}{ll}
        f_0^{(i)} & 1-f_1^{(i)} \\ 1-f_0^{(i)} & f_1^{(i)}
       \end{array} \right).
 \end{displaymath}
Here, $f_0^{(i)}$ is the probability by which a state $|0\rangle$ 
of the $i^{th}$ qubit is correctly identified as $|0\rangle$, while 
$1-f_1^{(i)}$ is the probability by which,
a state that is actually $|0\rangle$ is being wrongly considered as $|1\rangle$.
$f_0^{(i)}$ and $f_1^{(i)}$ are termed as the measurement fidelities
of the bases states $|0\rangle$ and $|1\rangle$ respectively of the $i^{th}$ qubit.
Using part of the measurement
data corresponding to the initial state preparation, we obtain the 
measurement fidelities:  $f_0^{(1)}=0.98$, $f_1^{(1)}=0.92$, $f_0^{(2)}=0.98$, 
$f_1^{(2)}=0.94$, $f_0^{(3)}=0.96$, $f_1^{(3)}=0.87$.
The intrinsic populations obtained in this manner are positive
(as observed till second decimal place). Using these intrinsic
population values, three-qubit density operators are obtained, 
that further undergo the convex optimization. The fidelities 
between the theoretically expected ($\rho_t$) and the experimentally
obtained ($\rho_e$) states is given by~\cite{Neill16}
\begin{equation}
 \mathcal{F}=Tr \sqrt{\sqrt{\rho_t}\rho_e\sqrt{\rho_t}}.
\end{equation}
These experimentally obtained three-qubit density operators are 
then used in our study to obtain the correlations, such as 
linear entropy of a single-qubit reduced state and a two-qubit 
entanglement measure, concurrence. We observe the variation of 
these correlations obtained from the experimental data with 
time and draw interesting observations, that are discussed in the maintext.

\begin{thebibliography}{10}

\bibitem{Billingsley}
Patrick Billingsley.
\newblock Prime numbers and brownian motion.
\newblock {\em The American Mathematical Monthly}, 80(10):1099--1115, 1973.

\bibitem{KusScharfHaake1987}
M~Ku{\'s}, R~Scharf, and F~Haake.
\newblock Symmetry versus degree of level repulsion for kicked quantum systems.
\newblock {\em Zeitschrift f{\"u}r Physik B Condensed Matter}, 66(1):129--134,
  1987.

\bibitem{Haake}
F.~Haake.
\newblock {\em Quantum Signatures of Chaos}.
\newblock Spring-Verlag, Berlin, 1991.

\bibitem{Peres02}
Asher Peres.
\newblock {\em Quantum Theory: Concepts and Methods}.
\newblock Kluwer Academic Publishers, New York, 2002.

\bibitem{CassidyEtal2009}
Amy~C. Cassidy, Douglas Mason, Vanja Dunjko, and Maxim Olshanii.
\newblock Threshold for chaos and thermalization in the one-dimensional
  mean-field bose-hubbard model.
\newblock {\em Phys. Rev. Lett.}, 102:025302, Jan 2009.

\bibitem{SantosRigol2010}
Lea~F. Santos and Marcos Rigol.
\newblock Onset of quantum chaos in one-dimensional bosonic and fermionic
  systems and its relation to thermalization.
\newblock {\em Phys. Rev. E}, 81:036206, Mar 2010.

\bibitem{Rigol16}
Luca D'Alessio, Yariv Kafri, Anatoli Polkovnikov, and Marcos Rigol.
\newblock From quantum chaos and eigenstate thermalization to statistical
  mechanics and thermodynamics.
\newblock {\em Advances in Physics}, 65(3):239--362, 2016.

\bibitem{JensenShankar1985}
R.~V. Jensen and R.~Shankar.
\newblock Statistical behavior in deterministic quantum systems with few
  degrees of freedom.
\newblock {\em Phys. Rev. Lett.}, 54:1879--1882, Apr 1985.

\bibitem{Deutsch91}
J.~M. Deutsch.
\newblock Quantum statistical mechanics in a closed system.
\newblock {\em Phys. Rev. A}, 43:2046--2049, 1991.

\bibitem{Srednicki94}
Mark Srednicki.
\newblock Chaos and quantum thermalization.
\newblock {\em Phys. Rev. E}, 50:888--901, 1994.

\bibitem{Rigol2009}
Marcos Rigol, Dunjko Vanja, and Olshanii Maxim.
\newblock Thermalization and its mechanism for generic isolated quantum
  systems.
\newblock {\em Nature}, 452(7189):854, 2009.

\bibitem{CiracHastings2011}
M.~C. Ba\~nuls, J.~I. Cirac, and M.~B. Hastings.
\newblock Strong and weak thermalization of infinite nonintegrable quantum
  systems.
\newblock {\em Phys. Rev. Lett.}, 106:050405, Feb 2011.

\bibitem{DeutchLiSharma2013}
J.~M. Deutsch, Haibin Li, and Auditya Sharma.
\newblock Microscopic origin of thermodynamic entropy in isolated systems.
\newblock {\em Phys. Rev. E}, 87:042135, Apr 2013.

\bibitem{LangenEtal2013}
T~Langen, R~Geiger, M~Kuhnert, B~Rauer, and J~Schmiedmayer.
\newblock Local emergence of thermal correlations in an isolated quantum
  many-body system.
\newblock {\em Nature}, 9:640, 2013.

\bibitem{LucaRigol2014}
Luca D'Alessio and Marcos Rigol.
\newblock Long-time behavior of isolated periodically driven interacting
  lattice systems.
\newblock {\em Phys. Rev. X}, 4:041048, Dec 2014.

\bibitem{LazaDasMoess2014}
Achilleas Lazarides, Arnab Das, and Roderich Moessner.
\newblock Periodic thermodynamics of isolated quantum systems.
\newblock {\em Phys. Rev. Lett.}, 112:150401, Apr 2014.

\bibitem{LazDasMoess2014pre}
Achilleas Lazarides, Arnab Das, and Roderich Moessner.
\newblock Equilibrium states of generic quantum systems subject to periodic
  driving.
\newblock {\em Phys. Rev. E}, 90:012110, Jul 2014.

\bibitem{Haldar2018}
Asmi Haldar, Roderich Moessner, and Arnab Das.
\newblock Onset of floquet thermalization.
\newblock {\em Phys. Rev. B}, 97:245122, Jun 2018.

\bibitem{Neill16}
C.~Neill, P.~Roushan, M.~Fang, Y.~Chen, M.~Kolodrubetz, Z.~Chen, A.~Megrant,
  R.~Barends, B.~Campbell, B.~Chiaro, A.~Dunsworth, E.~Jeffrey, J.~Kelly,
  J.~Mutus, P.~J.~J. O’Malley, C.~Quintana, D.~Sank, A.~Vainsencher,
  J.~Wenner, T.~C. White, A.~Polkovnikov, and J.~M. Martinis.
\newblock Ergodic dynamics and thermalization in an isolated quantum system.
\newblock {\em Nature Physics}, 12:1037, 2016.

\bibitem{Kaufman2016}
Adam~M. Kaufman, M.~Eric Tai, Alexander Lukin, Matthew Rispoli, Robert
  Schittko, Philipp~M. Preiss, and Markus Greiner.
\newblock Quantum thermalization through entanglement in an isolated many-body
  system.
\newblock {\em Science}, 353(6301):794--800, 2016.

\bibitem{ClosEtal2016}
Govinda Clos, Diego Porras, Ulrich Warring, and Tobias Schaetz.
\newblock Time-resolved observation of thermalization in an isolated quantum
  system.
\newblock {\em Phys. Rev. Lett.}, 117:170401, Oct 2016.

\bibitem{HazzardEtal2014}
Kaden R.~A. Hazzard, Mauritz van~den Worm, Michael Foss-Feig, Salvatore~R.
  Manmana, Emanuele~G. Dalla~Torre, Tilman Pfau, Michael Kastner, and Ana~Maria
  Rey.
\newblock Quantum correlations and entanglement in far-from-equilibrium spin
  systems.
\newblock {\em Phys. Rev. A}, 90:063622, Dec 2014.

\bibitem{Gutzwiller1990}
M.~C. Gutzwiller.
\newblock {\em Chaos in Classical and Quantum Mechanics}.
\newblock Springer-Verlag, New York, 1990.

\bibitem{Madhok2018_corr}
Vaibhav Madhok, Shruti Dogra, and Arul Lakshminarayan.
\newblock Quantum correlations as probes of chaos and ergodicity.
\newblock {\em Optics Communications}, 420:189 -- 193, 2018.

\bibitem{DavisHeller1981}
Michael~J. Davis and Eric~J. Heller.
\newblock Quantum dynamical tunneling in bound states.
\newblock {\em The Journal of Chemical Physics}, 75(1):246--254, 1981.

\bibitem{LinBallentine1990}
W.~A. Lin and L.~E. Ballentine.
\newblock Quantum tunneling and chaos in a driven anharmonic oscillator.
\newblock {\em Phys. Rev. Lett.}, 65:2927--2930, Dec 1990.

\bibitem{Peres1991}
Asher Peres.
\newblock Dynamical quasidegeneracies and quantum tunneling.
\newblock {\em Phys. Rev. Lett.}, 67:158--158, Jul 1991.

\bibitem{Tomsovic98b}
Steven Tomsovic, editor.
\newblock {\em Tunneling in complex systems}.
\newblock World Scientific, Singapore, 1998.

\bibitem{SrihariBook}
Srihari Keshavamurthy and Peter Schlagheck, editors.
\newblock {\em Dynamical Tunneling—Theory and Experiment}.
\newblock CRC Press, Boca Raton, FL, 2011.

\bibitem{Milburn99}
G.~J. Milburn.
\newblock Simulating nonlinear spin models in an ion trap, 1999.

\bibitem{Wang2004}
Xiaoguang Wang, Shohini Ghose, Barry~C. Sanders, and Bambi Hu.
\newblock Entanglement as a signature of quantum chaos.
\newblock {\em Phys. Rev. E}, 70:016217, 2004.

\bibitem{RuebeckArjendu2017}
Joshua~B. Ruebeck, Jie Lin, and Arjendu~K. Pattanayak.
\newblock Entanglement and its relationship to classical dynamics.
\newblock {\em Phys. Rev. E}, 95:062222, 2017.

\bibitem{Prosen2000}
Toma{\v{z}} Prosen.
\newblock Exact time-correlation functions of quantum ising chain in a kicking
  transversal magnetic fieldspectral analysis of the adjoint propagator in
  heisenberg picture.
\newblock {\em Progress of Theoretical Physics Supplement}, 139:191--203, 2000.

\bibitem{ArulSub2005}
Arul Lakshminarayan and V.~Subrahmanyam.
\newblock Multipartite entanglement in a one-dimensional time-dependent ising
  model.
\newblock {\em Phys. Rev. A}, 71:062334, Jun 2005.

\bibitem{Glauber}
Roy~J. Glauber and Fritz Haake.
\newblock Superradiant pulses and directed angular momentum states.
\newblock {\em Phys. Rev. A}, 13:357, Oct 1976.

\bibitem{Puri}
R.~R. Puri.
\newblock {\em Mathematical Methods of Quantum Optics}.
\newblock Springer, Berlin, 2001.

\bibitem{2018arXiv180600113S}
A.~{Seshadri}, V.~{Madhok}, and A.~{Lakshminarayan}.
\newblock {Tripartite mutual information, entanglement, and scrambling in
  permutation symmetric systems with an application to quantum chaos}.
\newblock {\em ArXiv e-prints}, May 2018.

\bibitem{SM}
See supplemental material at [url will be inserted by publisher] for further
  details.

\bibitem{Wootters}
William~K. Wootters.
\newblock Entanglement of formation of an arbitrary state of two qubits.
\newblock {\em Phys. Rev. Lett.}, 80:2245--2248, 1998.

\bibitem{YuEberly2007}
Ting Yu and J.~H. Eberly.
\newblock Evolution from entanglement to decoherence of bipartite mixed "x"
  states.
\newblock {\em Quantum Info. Comput.}, 7(5):459--468, July 2007.

\bibitem{GHZ0}
Daniel~M. Greenberger, Michael~A. Horne, and Anton Zeilinger.
\newblock Going beyond bell's theorem.
\newblock {\em arXiv:0712.0921 [quant-ph]}.

\bibitem{GHZ}
Daniel~M. Greenberger, Michael~A. Horne, Abner Shimony, and Anton Zeilinger.
\newblock Bell’s theorem without inequalities.
\newblock {\em American Journal of Physics}, 58(12):1131--1143, 1990.

\bibitem{Bhosale-pre-2017}
Udaysinh~T. Bhosale and M.~S. Santhanam.
\newblock Signatures of bifurcation on quantum correlations: Case of the
  quantum kicked top.
\newblock {\em Phys. Rev. E}, 95:012216, Jan 2017.

\bibitem{james-pra-2001}
Daniel F.~V. James, Paul~G. Kwiat, William~J. Munro, and Andrew~G. White.
\newblock Measurement of qubits.
\newblock {\em Phys. Rev. A}, 64:052312, 2001.

\bibitem{steffen-science-2006}
Matthias Steffen, M.~Ansmann, Radoslaw~C. Bialczak, N.~Katz, Erik Lucero,
  R.~McDermott, Matthew Neeley, E.~M. Weig, A.~N. Cleland, and John~M.
  Martinis.
\newblock Measurement of the entanglement of two superconducting qubits via
  state tomography.
\newblock {\em Science}, 313(5792):1423--1425, 2006.

\bibitem{lucero-prl-2008}
Erik Lucero, M.~Hofheinz, M.~Ansmann, Radoslaw~C. Bialczak, N.~Katz, Matthew
  Neeley, A.~D. O'Connell, H.~Wang, A.~N. Cleland, and John~M. Martinis.
\newblock High-fidelity gates in a single josephson qubit.
\newblock {\em Phys. Rev. Lett.}, 100:247001, 2008.

\end{thebibliography}
\end{document}